\documentclass[prd,twocolumn,preprintnumbers,superscriptaddress,nofootinbib]{revtex4}

\usepackage{mathtools}
\usepackage{xspace}
\usepackage{physics}
\usepackage{color}
\usepackage{units}
\usepackage{slashed}
\usepackage{hyperref}
\usepackage{cancel}
\usepackage{mathrsfs}
\usepackage[toc,page]{appendix}

\usepackage{graphicx}
\usepackage{epsfig}

\newcommand{\func}[3][1]{\if#11\mathinner{#2\!\left(#3\right)}\else\mathinner{#2^{#1}\!\left(#3\right)}\fi} 
\newcommand{\dfunc}[3][0]{\if#10\mathinner{#2\!\left(#3\right)}\else\mathinner{#2^{\left(#1\right)}\!\left(#3\right)}\fi} 
\newcommand{\yeq}{Y_{\mathrm{eq}}} 
\newcommand{\uI}{u^{\mathrm{I}}} 
\newcommand{\uII}{u^{\mathrm{II}}} 
\newcommand{\uIII}{u^{\mathrm{III}}} 
\newcommand{\xf}{x_{\mathrm{f}}} 
\newcommand{\vmoller}{v_{\text{M{\o}l}}} 
\newcommand{\mplanck}{M_{\mathrm{pl}}} 

\usepackage[utf8]{inputenc}
\DeclareUnicodeCharacter{2212}{-}

\begin{document}

\title{Asymptotic analysis of the Boltzmann equation for dark matter relic abundance}

\author{Logan A. Morrison}
\email{loanmorr@ucsc.edu}
\affiliation{Department of Physics and Santa Cruz Institute for Particle Physics,
University of California, Santa Cruz, CA 95064, USA}

\author{Hiren H. Patel}
\email{hpatel6@ucsc.edu}
\affiliation{Department of Physics and Santa Cruz Institute for Particle Physics,
University of California, Santa Cruz, CA 95064, USA}

\author{Jaryd F. Ulbricht}
\email{julbrich@ucsc.edu}
\affiliation{Department of Physics and Santa Cruz Institute for Particle Physics,
University of California, Santa Cruz, CA 95064, USA}

\hypersetup{
    pdftitle={Asymptotic analysis of the Boltzmann equation for dark matter relic abundance},
    pdfauthor={Logan A. Morrison, Hiren H. Patel, Jaryd F. Ulbricht}
}

\begin{abstract}
A solution to the Boltzmann equation governing the thermal relic abundance of cold dark matter is constructed by matched asymptotic approximations. The approximation of the relic density is an asymptotic series valid when the abundance does not deviate significantly from its equilibrium value until small temperatures. Resonance and threshold effects are taken into account at leading order and found to be negligible unless the annihilation cross section is negligible at threshold. Comparisons are made to previously attempted constructions and to the freeze out approximation commonly employed in the literature. Extensions to higher order matching is outlined, and implications for solving related systems are discussed. We compare our results to a numerical determination of the relic abundance using a benchmark model and find a fantastic agreement. The method developed also serves as a solution to a wide class of problems containing an infinite order turning point. 
\end{abstract}

\maketitle

\section{Introduction} \label{sec:intro}

Successful cosmological theories must explain various observations, such as the structure of the cosmic microwave background, baryonic acoustic oscillations, structure formation, among others. These observations always require a cold, nearly electromagnetically-neutral, non-baryonic matter component, which we call dark matter (DM)~\cite{particle2006review,profumo2017}. Observations with Plank~\cite{planck2018} show that almost \(2/5\) of all matter in the Universe consists of DM. While we know the basic properties of DM (it interacts gravitationally and at most weakly with the known Standard Model (SM) particles), the precise nature of DM is unknown. Some of the most popular theories of DM involve extending the SM by adding new particles. DM candidates often arise naturally in models trying to address other outstanding issues such as the hierarchy problem, the strong CP problem, and neutrino masses (e.g., neutralinos in super-symmetry~\cite{jungman1996supersymmetric}, axions~\cite{duffy2009axions} and sterile neutrinos~\cite{boyarsky2019sterile}, respectively.)

For any theory of particle DM to be viable, the theory must produce DM with the observed relic abundance of $\Omega_{\mathrm{DM}}h^2 \equiv \rho_{\mathrm{DM}}h^2 / \rho_{\mathrm{crit}}\approx 0.12$~\cite{planck2018}, where the relative uncertainty of today's value of Hubble's parameter \(H_{0}\) is absorbed into the dimensionless Hubble parameter \(h\)
\begin{equation}
    H_{0} \equiv h \times 100 \; \mathrm{km} \; \mathrm{s}^{-1} \; \mathrm{Mpc}^{-1}.
\end{equation}
Therefore, it is necessary to be able to compute the abundance of DM for a given theory accurately. The standard method for determining the abundances of DM for a given theory is by solving the Boltzmann equation, which in the standard Friedman-Roberston-Walker cosmology is~\cite{Gondolo1991}:
\begin{align}
    \pdv{f_{\chi}}{t} - H \dfrac{|\vb{p}|^2}{E}\pdv{f_{\chi}}{E} = \mathcal{C}[f_{\chi}],
\end{align}
where $f_{\chi}(\vb{p},t)$ is the DM phase-space distribution, $\vb{p}$ the DM momentum, $E=\qty(\vb{p}^2 + m_{\chi}^2)^{1/2}$, $H$ the Hubble scale and $\mathcal{C}[f]$ the collision term which depends on the details of the DM model. In all but a select few cases it is sufficient to take the first momentum of this equation\footnote{See Ref.~\cite{binder2017early} for an example where more than just the first moment of the Boltzmann equation must be considered.}, which, in the cases where the DM interacts with the SM via $\chi\bar{\chi}\leftrightarrow \mathrm{SM}$, takes the form of:
\begin{equation}
    \dv{n_{\chi}}{t} + 3Hn_{\chi} = -\expval{\sigma_{\chi\bar{\chi}\to\mathrm{SM}} \vmoller}\qty(n_{\chi}^2 - n_{\chi,\mathrm{eq}}^2),
\end{equation}
where \(n_{\chi}\) is the DM number density
\begin{equation}
    n_{\chi} = \int\frac{\dd[3]{\vb{p}}}{(2\pi)^3}f_{\chi},
\end{equation}
\(n_{\chi, \mathrm{eq}}\) is the DM equilibrium number density obtained by setting $f_{\chi}=f_{\chi,\mathrm{eq}}$ given by a Bose-Einstein or Fermi-Dirac distribution: $1/\qty[\exp(E/T) \pm 1]$ depending on the statistics of the DM particle, and $\expval{\sigma_{\chi\bar{\chi}\to\mathrm{SM}} \vmoller}$ (which we will shorten to $\expval{\sigma\vmoller}$) is the thermally-averaged cross section:
\begin{align}
    \expval{\sigma\vmoller} =\dfrac{
    \int\sigma\vmoller f_{\chi,\mathrm{eq}}(E_1)f_{\chi,\mathrm{eq}}(E_2)\dd[3]{\vb{p}_1}\dd[3]{\vb{p}_2}
    }{
    \int f_{\chi,\mathrm{eq}}(E_1)f_{\chi,\mathrm{eq}}(E_2)\dd[3]{\vb{p}_1}\dd[3]{\vb{p}_2}
    },
\end{align}
with $\sigma$ being the zero-temperature cross section for $\chi\bar{\chi}\to\mathrm{SM}$. This form of the Boltzmann equation is often modified to absorb the effects of the of the expanding Universe by scaling the solutions with the entropy density of the SM, \(s\), through $Y\equiv n_{\chi} / s$. We then have the following differential equation:
\begin{equation}\label{eq:lee-weinberg}
\begin{gathered}
\dv{\func{Y}{x}}{x} = -\lambda \func{f}{x} \qty[ \func[2]{Y}{x} - \func[2]{\yeq}{x} ]\,,\\
\end{gathered}
\end{equation}
The dependent variable \(Y\) is the comoving number density of a particle species (it is common to refer to \(Y\) as the \textit{abundance} for brevity), i.e. the number of particles per cosmic comoving volume element. The independent variable \(x \equiv m_{\chi} / T\) is the ratio of the particle mass to the temperature of the thermal bath. The equilibrium abundance, \(\yeq\), is the comoving number density of a particle species when in thermal (chemical) equilibrium with the thermal bath. The prefactor \(\lambda f\) contains the cross section of the particle species and is given by
\begin{equation} \label{eq:lambdaf}
    \lambda \func{f}{x} \equiv \sqrt{\frac{\pi}{45}} \frac{m \mplanck}{x^{2}} \sqrt{\func{g_{*, \mathrm{eff}}}{x}} \expval{\sigma \vmoller},
\end{equation}
where \(\expval{\sigma \vmoller}\) implicitly depends on \(x\) and \(\func{g_{*, \mathrm{eff}}}{x}\) is a function characterising the effective number of degrees of freedom contributing to the energy density and entropy density of the universe:
\begin{equation}
    \sqrt{\func{g_{*, \mathrm{eff}}}{x}} \equiv \frac{\func{h_{\mathrm{eff}}}{x}}{\sqrt{\func{g_{\mathrm{eff}}}{x}}} \qty(1 + \frac{1}{3} \dv{\ln(\func{h_{\mathrm{eff}}}{x})}{\ln(x)}).
\end{equation}
The effective number of degrees of freedom contributing to the total energy density and entropy density are \(\func{g_{\mathrm{eff}}}{x}\) and \(\func{h_{\mathrm{eff}}}{x}\) respectively. The limiting behavior \(Y_{\infty} \equiv \lim_{x\to\infty} \func{Y}{x}\) of the solution is the quantity of interest, and determines the thermal relic density.

The starting point for our analysis is Equation~\eqref{eq:lee-weinberg}, but it cannot be solved exactly, and therefore one resorts to obtaining approximations.  The most common method of approximation is direct numerical integration.  The use of general-purpose integrators tend to fail due to the largeness of $\lambda$, and even sophisticated algorithms like Radau5~\cite{hairer1996}, LSODA~\cite{lsoda} struggle because the differential equation is exceptionally stiff which requires high precision arithmetic.  Dedicated software packages to obtain dark matter relic abundances from particle physics models such as \texttt{micrOMEGAS}~\cite{BELANGER2014960}, and DarkSUSY~\cite{Bringmann_2018}, etc. fare better due to additional heuristics supplied to their integrators.  However, these canned software packages designed to solve \eqref{eq:lee-weinberg} are compatible with only a small subset of beyond-standard-model (BSM) scenarios, which limits the end user from performing an analysis of more exotic models such as those with Lorentz violation or large \(\mathcal{N}\) Yang-Mills~\cite{Morrison2020}.

An alternative approach to obtaining the limiting behavior of \eqref{eq:lee-weinberg} is to look for analytic approximations.  Several approximations exist in the literature such as ~\cite{Gondolo1991, kolb1994early}, and can provide results accurate to 1-5\%, confirmed by comparing against results of numerical integration. However, by nature of their construction it is not possible to systematically improve upon these approximations simply because there is no way to assign a parametric dependence on the error.

The mathematical technique allowing for the construction of approximations while bounding the error is \textit{asymptotic analysis} (for an in depth review of perturbation theory and asymptotics see~\cite{bender1999advanced}). The error is managed by a controlling parameter such that, as the controlling parameter is taken arbitrarily close to some limit point, the error vanishes relative to the approximation. It is in this sense that we can consider the error to be `small'. 
A natural choice for the problem at hand is to choose $\lambda$ in \eqref{eq:lee-weinberg} as the controlling parameter, and to attempt to construct an asymptotic approximation in the limit $\lambda\rightarrow \infty$.


The authors of \cite{Bender2012} attempted to construct an asymptotic approximation by using boundary-layer-analysis, yielding a technically more correct result with the requisite scaling behavior of the error. However, we found their their matching procedure to be inconsistent.  We were able to correct these errors to arrive at similar results.  But in order to get a good approximation we had to perform a resummation of the largest terms of a divergent series, and for this reason we found it more intuitive to take a different approach, based on the Wentzel–Kramers–Brillouin (WKB) technique.

In this paper, we present our asymptotic approximation to \eqref{eq:lee-weinberg}. Our final results are given by \eqref{eq:results:freeze-out}, \eqref{eq:results:Yinf}, and \eqref{eq:results:deltax}. The paper is structured as follows: In Section \ref{sec:behaviors} we derive the large and small \(x\) behavior of the solution as well as the large \(x\) asymptotic behavior of the thermal cross section and equilibrium abundance for later reference. To \eqref{eq:lee-weinberg} we associate a second order linear differential equation of Schr\"odinger type, making a WKB analysis possible. However, there exists an infinite order turning point (where the potential and all its derivatives vanish) at \(x = \infty\). Such classes of differential equations are notoriously difficult to solve, so to circumvent this issue we employ a more robust \textit{uniform} WKB ansatz in Section \ref{sec:solutions} that is better suited to the infinite order turning point problem, and construct asymptotic solutions in three subregions of \(x \in (0,\infty )\): The thermal equilibrium region (I), freeze-out region (II), and post-freeze-out region (III). We preform an asymptotic match of region I and III in Section \ref{sec:matching} at leading and next to leading order, removing all undetermined constants. After matching we take the limit \(x\to\infty\), yielding an asymptotic approximation of the relic density. In Section \ref{sec:results} we collect our results and compare our approximation against a numerical determination of the relic density using a benchmark model. We find that our approximation, when compared to numerical results, gives sub-percent errors when the dark matter candidate freezes out at roughly \(x = 25\). To our knowledge, we are the first to present an asymptotic approximation to \(Y_{\infty}\). We are also unaware of a previous application of this method to the infinite order turning point problem.

\section{Asymptotic behaviors} \label{sec:behaviors}

We briefly discuss the asymptotic behavior of some of the quantities in \eqref{eq:lee-weinberg} and the general large and small \(x\) behavior of the solution. The equilibrium abundance of a particle species is given by
\begin{equation}
    \func{\yeq}{x} = A \int^{\infty}_{0} \frac{s^{2} \dd{s}}{e^{\sqrt{s^{2} + x^{2}}} \mp 1},
\end{equation}
where the upper sign is for bosons and the lower for fermions, and \(A\) is given by
\begin{equation}
    A \equiv \frac{45}{4 \pi^{4}} \frac{g}{\func{h_{\mathrm{eff}}}{x}},
\end{equation}
where \(g\) is the number of internal degrees of freedom of the particle species and \(\func{h_{\mathrm{eff}}}{x}\) is the number of relativistic degrees of freedom contributing to the entropy density. The large $x$ behavior of the the equilibrium abundance is
\begin{equation}
    \func{\yeq}{x} \sim \sqrt{\frac{\pi}{2}} A x^{3/2} e^{-x}, \quad \qty(x \to \infty).
\end{equation}

\begin{figure}[ht!]
\includegraphics[width=\columnwidth]{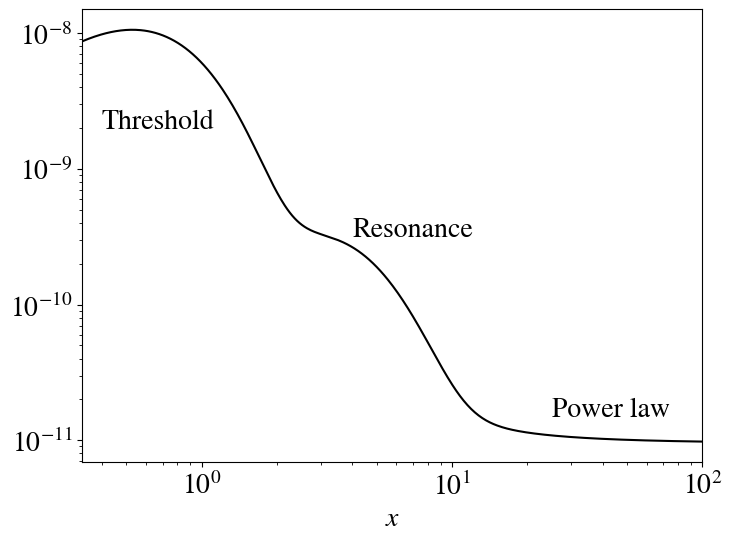}
\caption{Typical behavior of the thermally averaged annihilation cross section \(\expval{\sigma \vmoller}\) as a function of \(x\), the vertical axis is \(\text{GeV}^{-2}\). On the left threshold effects dominate. In the center threshold effects have decayed away and resonance contributions now dominate. On the right all threshold and resonance effects are negligible and the thermally averaged cross section assumes the form of a power law (in this case a constant).\label{fig:thermal_xsec}}
\end{figure}

For temperatures \(T \lesssim 3 m\), where \(m\) is the mass of particle species in question, the phase space distribution for all statistics is well approximated by the Maxwell-Boltzmann distribution. Making this substitution the \(2 \to \text{all}\) thermally averaged annihilation cross section reduces to a single integral~\cite{Gondolo1991},
\begin{subequations}
\begin{align} \label{eq:thermalxsec}
\expval{\sigma \vmoller} \sim& \; \int^{\infty}_{0} \dd{\epsilon} \sigma v_{\mathrm{lab}} \func{\mathscr{K}}{\epsilon, x}, \quad \qty(x \to \infty),\\[2mm] \label{eq:thermalkernel}
\func{\mathscr{K}}{\epsilon, x} =& \; \frac{2 x}{\func[2]{K_{2}}{x}} \epsilon^{1/2} \qty(1 + 2 \epsilon) \func{K_{1}}{2 x \sqrt{1 + \epsilon}},
\end{align}
\end{subequations}
where
\begin{equation}
    \epsilon \equiv \frac{s - 4 m^{2}}{4 m^{2}}.
\end{equation}
We can further approximate the thermal kernel \eqref{eq:thermalkernel} using the large argument expansion of the modified Bessel function.
\begin{subequations}
\begin{align} \label{eq:thermalkernel_asymp}
    \func{\mathscr{K}}{\epsilon, x} \sim& \; \frac{2 x^{3/2}}{\sqrt{\pi}} \frac{\epsilon^{1/2} \qty(1 + 2 \epsilon)}{\qty(1 + \epsilon)^{1/4}} e^{x \phi \qty(\epsilon)}, \quad \qty(x \to \infty),\\[2mm]
    \func{\phi}{\epsilon} =& - 2 \qty(\sqrt{1 + \epsilon} - 1),
\end{align}
\end{subequations}
When \(x\) is very large we can estimate the integral using Laplace's method. We first located the maximum of the integrand in \eqref{eq:thermalxsec}, and denote this point \(\epsilon_{0}\). In the limit that \(x \to \infty\) this maximum is just the maximum of \(\func{\phi}{\epsilon}\). The approximation of the thermally averaged cross section then has a residual exponential character \(\exp[x \func{\phi}{\epsilon_{0}}]\). If \(\sigma v_{\mathrm{lab}}\) is sufficiently smooth, i.e. any resonances are broad and all annihilation channels are of similar scale, then \(\epsilon_{0} = 0\) and \(\func{\phi}{0} = 0\), so the thermally averaged cross section goes like some power of \(x\). In this case a more thorough treatment, using Watson's lemma, yields:
\begin{equation} \label{eq:thermalxsecseries}
    \expval{\sigma \vmoller} \sim \sum^{\infty}_{k = 0} \sigma_{k} x^{-k}, \quad \qty(x \to \infty),
\end{equation}
where the coefficients \(\sigma_{k}\) are easily found. There are two common scenarios in which the estimate \eqref{eq:thermalxsecseries} breaks down for intermediate values of \(x\): when the annihilation cross section contains a very narrow resonance or the dominant annihilation channel has support only when \(s > 4 m^{2}\). For a narrow resonance the annihilation cross section approaches a delta function in the limit that the width of the resonance goes to 0. If this narrow resonance is centered at \(s = m^{2}_{\mathrm{R}}\) then \(\epsilon_{0} \sim \qty(m^{2}_{R} - 4 m^{2}) / 4 m^{2}\) and
\begin{equation*}
    \func{\phi}{\epsilon_{\mathrm{0}}} \sim - \frac{m_{\mathrm{R}} - 2 m}{m}.
\end{equation*}
Alternatively, if there exists an annihilation channel that is kinematically unavailable when \(s < s_{\mathrm{Th}} = 4 m^{2}_{\mathrm{Th}}\), but that dominates the cross section when \(s > s_{\mathrm{Th}}\), then \(\epsilon_{0} \sim \qty(m^{2}_{\mathrm{Th}} - m^{2}) / m^{2}\) and
\begin{equation*}
    \func{\phi}{\epsilon_{0}} \sim - \frac{2 \qty(m_{\mathrm{Th}} - m)}{m}.
\end{equation*}
Therefore, we can characterize the thermally averaged cross section for intermediate to large \(x\) by
\begin{subequations}
\begin{align} \label{eq:sigmav_largex_general}
\expval{\sigma \vmoller} \sim& x^{\beta} e^{- \alpha x} \sum^{\infty}_{k = 0} c_{k} x^{-k}, \quad \qty(x \to \infty)\\[2mm]
\alpha =& \begin{cases}
0 & \text{Power Law}\\[2mm]
\frac{m_{\mathrm{R}} - 2 m}{m} & \text{Resonance}\\[2mm]
\frac{2 \qty(m_{\mathrm{Th}} - m)}{m} & \text{Threshold}
\end{cases} \label{eq:alpha}
\end{align}
\end{subequations}
The coefficients \(\beta\) and \(c_{k}\) generally depend on the choice of \(\alpha\). It is almost always the case that the leading order behavior of the thermally averaged cross section has no exponential decay (i.e. \(\alpha = 0\)) for very large \(x\). We then expect that \(\alpha\) will make rapid transitions as we move from intermediate \(x\) to large \(x\), ultimately going to 0 once \(x\) becomes sufficiently large. This prediction is validated in Fig.~\ref{fig:thermal_xsec}. Substituting the power law approximation \eqref{eq:thermalxsecseries} into \eqref{eq:lambdaf} yields the standard behavior of \(\func{f}{x}\) for large \(x\),
\begin{equation} \label{eq:fLO}
    \func{f}{x} \sim x^{- n - 2}, \quad \qty(x \to \infty),
\end{equation}
where \(n\) is the order of the first non vanishing term in \eqref{eq:thermalxsecseries}. The normalization of the thermally averaged cross section has been stripped away and included in the parameter \(\lambda\).

For \(x\) not too large we approximate the solution to \eqref{eq:lee-weinberg} by assuming a formal series expansion in powers of \(1/\lambda\):
\begin{equation} \label{eq:Y_large_lambda_sum}
    \func{Y}{x} \sim \sum^{\infty}_{k = 0} \func{Y_{k}}{x} \lambda^{-k}, \quad \qty(\lambda \to \infty).
\end{equation}
This gives the approximate solution
\begin{equation} \label{eq:Ysmallx}
    \func{Y}{x} \sim \func{\yeq}{x} - \frac{\func{\yeq'}{x}}{2 \lambda \func{f}{x} \func{\yeq}{x}}, \quad \qty(\lambda \to \infty).
\end{equation}
Because \(\yeq\) decays exponentially fast this solution becomes invalid when \(\lambda \func{f}{x} \func{\yeq}{x} = \order{1}\). When \(x\) is very large, such that \(\lambda \func{f}{x} \func{\yeq}{x} \ll 1\) we can neglect the last term on the right hand side of \eqref{eq:lee-weinberg}, resulting in a second approximation
\begin{equation} \label{eq:Ylargex}
    \func{Y}{x} \sim \qty[\frac{1}{Y_{\infty}} - \lambda \int^{\infty}_{x} \func{f}{s} \dd{s}]^{-1}, \quad \qty(x \to \infty). 
\end{equation}
Assuming \(Y_{\infty} > 0\), and because the integral
\begin{equation*}
    \lambda \int^{\infty}_{x} \func{f}{s} \dd{s}
\end{equation*}
generally diverges as \(x \to 0\), there necessarily exists some \(0 < x_{\mathrm{pole}} < \infty\) such that
\begin{equation*}
    \frac{1}{Y_{\infty}} - \lambda \int^{\infty}_{x_{\mathrm{pole}}} \func{f}{s} \dd{s} = 0
\end{equation*}
This approximation is therefore only valid when \(x \gg x_{\mathrm{pole}} > 0\), and we cannot satisfy the boundary condition at \(x = 0\). 

\begin{figure}[ht!]
\centering
\includegraphics[width=\columnwidth]{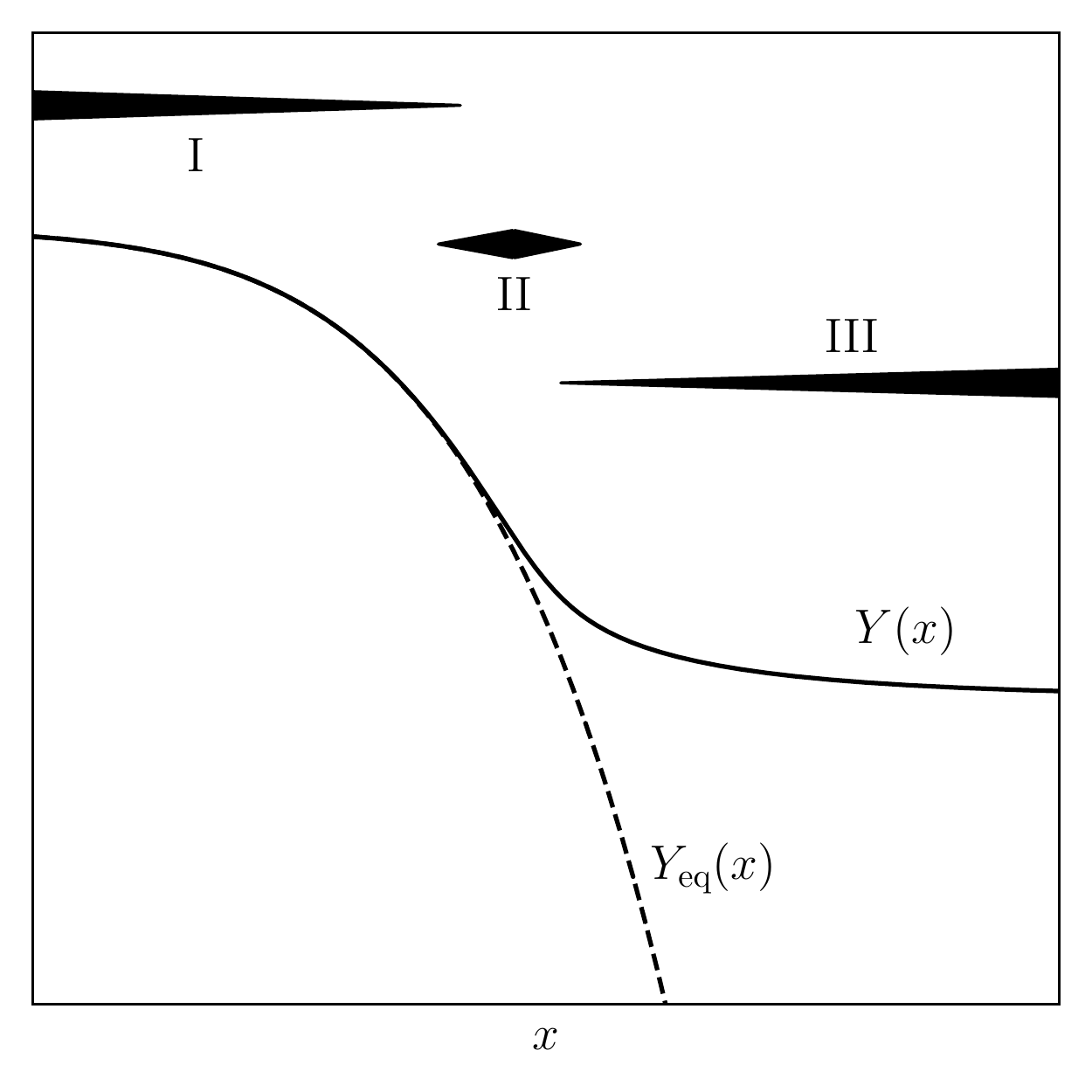}
\caption{The abundance of a particle species as a function of \(x \equiv m / T\). In region I the abundance closely tracks the equilibrium abundance. In region III the abundance asymptotes to a constant, denoted \(Y_{\infty}\). Region II represents the transition between region I and III. The wedges sketch the extent of each of the regions.}
\label{fig:schematic}
\end{figure}

Generally the approximate solutions \eqref{eq:Ysmallx} and \eqref{eq:Ylargex} have no overlap in their region of validity, so the arbitrary constant \(Y_{\infty}\) cannot yet be determined. One must either construct an intermediate solution whose region of validity overlaps with both the large \(\lambda\) and the large \(x\) approximations, or modify one or both solutions such that they have some overlap. We take the latter approach, essentially resumming the largest components of \eqref{eq:Y_large_lambda_sum} to all orders.

\section{Construction of asymptotic solutions} \label{sec:solutions}

We begin by transforming \eqref{eq:lee-weinberg} from a first order non-linear differential (Riccati) equation into a second order linear differential equation of the Schr\"odinger type by changing the dependent variable to
\begin{equation} \label{eq:y2u}
\func{Y}{x} = \frac{1}{\lambda \func{f}{x}} \dv{x} \ln(\sqrt{\lambda \func{f}{x}} \func{u}{x})\,,
\end{equation}
so that
\begin{equation}
\begin{gathered}\label{eq:udiffeq}
u'' - \Big[(\lambda f Y_\text{eq})^2 + \frac{3}{4}\Big(\frac{f'}{f}\Big)^2 - \frac{1}{2}\frac{f''}{f}\Big] u = 0\,,\\
\frac{u'(0)}{u(0)} = \lambda \mathinner{f(0)} \mathinner{Y_\text{eq}(0)} - \frac{f'(0)}{2 f(0)}
\end{gathered}
\end{equation}
Using the canonical WKB ansatz,
\begin{equation}
\func{u}{x} \sim \exp(\lambda \sum^{\infty}_{k=0} \func{S_{k}}{x} \lambda^{-k}), \quad \qty(\lambda \to \infty),    
\end{equation}
gives the solution for \(Y\) as a formal power series in \(1/\lambda\).
\begin{equation}
    \func{Y}{x} \sim \sum^{\infty}_{k=0} \frac{\func{S'_{k}}{x}}{\func{f}{x}} \lambda^{-k} + \frac{\func{f'}{x}}{2 \lambda \func[2]{f}{x}}, \quad \qty(\lambda \to \infty).
\end{equation}
We see that the \(1/\lambda\) series solution of \eqref{eq:lee-weinberg} is equivalent to the WKB solution of \eqref{eq:udiffeq}.

In what follows we construct asymptotic approximations for small \(x\) (Region I), large \(x\) (Region III), and intermediate \(x\) (Region II), shown schematically in Fig.~\ref{fig:schematic}. The region II approximation is superfluous, as we will see the domain of validity of the region I and III solutions generally overlap (and hence the region I approximation can be asymptotically matched directly onto the region III approximation). However, the approximation in the overlap region motivates a definition of a freeze-out temperature that ensures a consistent asymptotic expansion in all three regions. In order to simplify our notation we define:
\begin{subequations}
\begin{align}
    \label{eq:Q}
    \func{Q}{x} \equiv& \func{f}{x} \func{\yeq}{x},\\[2mm]
    \label{eq:P}
    \func{P}{x} \equiv& \frac{3}{4}\qty(\frac{\func{f'}{x}}{\func{f}{x}})^2 - \frac{1}{2}\frac{\func{f''}{x}}{\func{f}{x}},
\end{align}
\end{subequations}
so that \eqref{eq:udiffeq} becomes
\begin{equation}\label{eq:udiffeq2}
u'' - \qty[\lambda^{2} \func[2]{Q}{x} + \func{P}{x}] u = 0.
\end{equation}

\begin{figure}[t]
    \centering
    \includegraphics[width=\columnwidth]{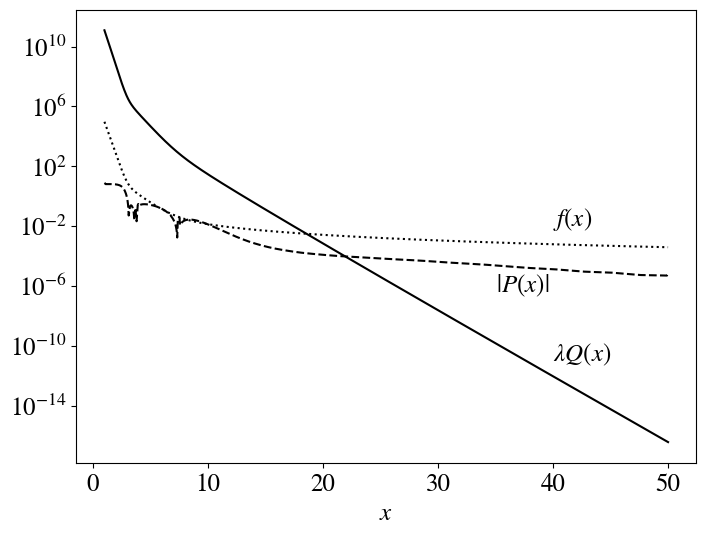}
    \caption{The functions \(\func{f}{x}\), \(\func{P}{x}\), and \(\lambda \func{Q}{x}\) using a benchmark model (see Section~\ref{sec:results}).\label{fig:fPQ}}
\end{figure}

Before proceeding we make some observations about the behavior of these two functions \(\func{Q}{x}\) and \(\func{P}{x}\). Consider, for example, the following large \(x\) behavior of \(f\) from \eqref{eq:sigmav_largex_general}: 
\begin{equation}
    \func{f}{x} \sim x^{\beta} e^{-\alpha x}, \quad \qty(x \to \infty).
\end{equation}
The resulting behavior for \(\func{Q}{x}\) and \(\func{P}{x}\) is 
\begin{subequations}
\begin{align}
    \func{Q}{x} \sim& \sqrt{\frac{\pi}{2}} A x^{\beta + 3/2} e^{- \qty(1 + \alpha) x}, \quad \qty(x \to \infty), \label{eq:Qlargex}\\[2mm]
    \func{P}{x} \sim& \frac{\alpha^{2}}{4} - \frac{\alpha \beta}{2 x} + \frac{\beta \qty(2 + \beta)}{2 x^{2}}, \quad \qty(x \to \infty). \label{eq:Plargex}
\end{align}
\end{subequations}
and are shown in Fig.~\ref{fig:fPQ}.  Because of the exponential decay in \eqref{eq:Qlargex} \(\func{Q}{x}\) and all its derivatives vanish as \(x \to \infty\). Note that there are two linearly independent solutions to \eqref{eq:udiffeq2}, and the WKB approximations of these two solutions are multivalued. Therefore, if we approximate the full solution as a specific combination of these two linearly independent solutions near \(x = \infty\), the same combination cannot be used for \(x e^{2 \pi i}\). This is known as the Stoke's phenomenon. Essentially, the problem is that the approximations are necessarily domain dependent. In this case, because \(x = \infty\) is an essential singularity, in the neighborhood of the turning point there exists an infinite number of domains (bounded by Stoke's and Anti-Stokes lines), each requiring a different combination of linearly independent solutions. This is the infinite order turning point problem.

\subsection{Thermal equilibrium region (Region I)} \label{subsec:regionI}

In the small \(x\) region, we construct a uniform WKB approximation to \eqref{eq:udiffeq}.  The ansatz, originally constructed by Langer~\cite{Langer1937}, is:
\begin{equation}\label{eq:reg1-ansatz}
\func{\uI}{x} =i \qty(\frac{S}{S'})^{1/2} \func{K_{\nu}}{\lambda S},
\end{equation}
where \(\func{K_{\nu}}{z}\) is the modified Bessel function of the second kind, and \(\func{S}{x}\) admits a series expansion in \(1/\lambda\),
\begin{equation}\label{eq:serS}
\func{S}{x} \sim \sum^{\infty}_{k=0} \func{S_{k}}{x} \lambda^{-2k}, \quad \qty(\lambda \to \infty).
\end{equation}
The order parameter \(\nu\) is left to be determined, it will be chosen to prolong the validity of the approximation. We remark that with the choice \(\nu = \tfrac{1}{2}\) the ansatz \eqref{eq:reg1-ansatz} reduces to standard WKB. This construction is particularly well suited to the infinite order turning point problem, as can be seen by considering the exact solutions of
\begin{equation}
    \dv[2]{y}{t} - \qty[\lambda^{2} e^{- 2 t} + \nu^{2}] y = 0,
\end{equation}
which are linear combinations of modified Bessel functions \(\func{I_{\nu}}{\lambda e^{- t}}\), \(\func{K_{\nu}}{\lambda e^{-t}}\).

The prefactor of \eqref{eq:reg1-ansatz} is chosen so that \eqref{eq:udiffeq} becomes a differential equation entirely in terms of \(\func{S}{x}\), and the factor of \(i\) ensure the solution is real-valued for positive \(x\). Substituting the ansatz \eqref{eq:reg1-ansatz} into \eqref{eq:udiffeq2} and then inserting \eqref{eq:serS} into the resulting equation allows one to solve for each term by equating powers of \(1/\lambda\):
\begin{widetext}
\begin{equation} \label{eq:Sdiffeq}
\begin{gathered}
    - \frac{1}{2} \qty(\frac{S'''}{S'}) + \frac{3}{4} \qty(\frac{S''}{S'})^{2} + \qty(\nu^{2} - \frac{1}{4}) \qty(\frac{S'}{S})^{2} + \lambda^{2} \qty(S')^{2} - \lambda^{2} \func[2]{Q}{x} - \func{P}{x} = 0.\\[2mm]
    \lambda \func{Q}{0} - \frac{1}{2} \frac{\func{f'}{0}}{\func{f}{0}} = - \lambda \func{S'}{0} \frac{\func{K_{\nu + 1}}{\lambda \func{S}{0}}}{\func{K_{\nu}}{\lambda \func{S}{0}}} + \frac{1}{2} \qty[\qty(1 + 2 \nu) \frac{\func{S'}{0}}{\func{S}{0}} - \frac{\func{S''}{0}}{\func{S'}{0}}]
\end{gathered}
\end{equation}
\end{widetext}
Solving \eqref{eq:Sdiffeq} at leading order gives
\begin{equation}
    \func{S_{0}}{x} = - \int^{x}_{0} \func{Q}{s} \dd{s} + \func{S_{0}}{0}.
\end{equation}
The boundary condition dictates that the sign of \(\func{S'_{0}}{x}\) must be negative, but the initial value is arbitrary. If the leading order solution changes sign at some finite value of \(x\) we will have to contend with the Stoke's phenomenon, so we require \(\func{S_{0}}{x}\) be bounded from below. This is guaranteed with the choice \(\func{S_{0}}{\infty} = 0\), yielding
\begin{equation} \label{eq:S0}
    \func{S_{0}}{x} = \int^{\infty}_{x} \func{Q}{s} \dd{s}.
\end{equation}

We now estimate \(\func{S_{0}}{x}\) for large \(x\) . Begin by making a change of variables to \(t \equiv s / x\).
\begin{equation} \label{eq:S0changeofvariables}
    \func{S_{0}}{x} = x \int^{\infty}_{1} \func{Q}{xt} \dd{t}
\end{equation}
If \(x\) is large \(\func{Q}{x t}\) is exponentially suppressed everywhere along the range of integration. Then write
\begin{equation} \label{eq:QwF}
    \func{Q}{x} \sim \func{F}{x} e^{- \qty(1 + \alpha) x}, \quad \qty(x \to \infty),
\end{equation}
where we assume that \(\func{F}{x}\) contains no exponential terms. If resonance or threshold effects are negligible we will set \(\alpha = 0\). Inserting \eqref{eq:QwF} into \eqref{eq:S0changeofvariables} and expanding \(\func{F}{xt}\) as a Taylor series around \(x\) then gives
\begin{equation} \label{eq:S0series}
    \func{S_{0}}{x} \sim \frac{\func{Q}{x}}{\func{F}{x}} \sum^{\infty}_{n = 0} \frac{1}{\qty(1 + \alpha)^{n+1}} \dv[n]{\func{F}{x}}{x}, \;\, \qty(x \to \infty).
\end{equation}
The errors introduced are exponentially small as \(x \to \infty\). Because \(\func{F}{x}\) contains no exponential terms by assumption this series naturally organises itself as an expansion in powers of \(1/x\).

\begin{figure}[t]
\includegraphics[width=\columnwidth]{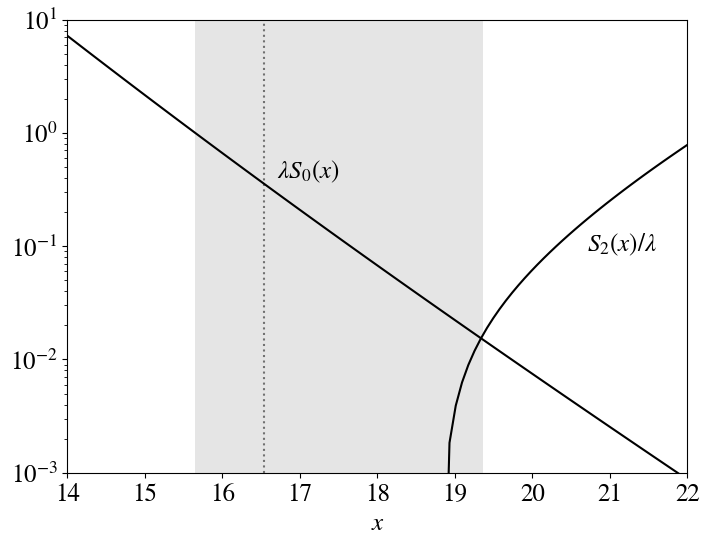}
\caption{The leading order term \(\lambda \func{S_{0}}{x}\) compared to the next to leading order term \(\func{S_{2}}{x} / \lambda\). The shaded region corresponds to where \(1 \geq \lambda S_{0} \geq S_{2} / \lambda\). This shaded region also represents the extension of the region of validity over standard WKB. The vertical dotted line indicates the location of what will eventually be defined as the freeze-out temperature.\label{fig:S0S2}}
\end{figure}

Solving for the next to leading order term in \eqref{eq:Sdiffeq} we find
\begin{widetext}
\begin{equation} \label{eq:S2}
    \func{S_{2}}{x} = \frac{\qty(4 \nu^{2} - 1)}{8} \qty[\frac{1}{\func{S_{0}}{x}} - \frac{1}{\func{S_{0}}{0}}] - \underbrace{\frac{1}{2} \int^{x}_{0} \frac{\dd{s}}{\func{Q}{s}} \qty{\func{P}{s} + \frac{1}{2} \frac{\func{Q''}{s}}{\func{Q}{s}} - \frac{3}{4} \qty[\frac{\func{Q'}{s}}{\func{Q}{s}}]^{2}} + \func{S_{2}}{0}}_{\text{Standard WKB}}.
\end{equation}
\end{widetext}
The integral can be approximated in a very similar way as for \(\func{S_{0}}{x}\). We report only the leading order term:
\begin{equation}
    \func{S_{2}}{x} \sim - \frac{1}{2 \qty(1 + \alpha)} \frac{\func{\phi}{x}}{\func{Q}{x}}, \quad \qty(x \to \infty),
\end{equation}
where
\begin{equation} \label{eq:phi}
    \func{\phi}{x} \equiv P - \nu^{2} \qty(\frac{S'_{0}}{S_{0}})^{2} - \sqrt{\frac{S'_{0}}{S_{0}}} \dv[2]{x} \sqrt{\frac{S_{0}}{S'_{0}}}.
\end{equation}
\(\func{S_{2}}{x}\) is then exponentially increasing as \(x \to \infty\). In order to extend the region of validity of our approximation we choose \(\nu\) to cancel the leading order large \(x\) component of \eqref{eq:phi}. The last term of \eqref{eq:phi} is at most of order \(\order{1/x^{3}}\),
\begin{equation}
    \sqrt{\frac{S'_{0}}{S_{0}}} \dv[2]{x} \sqrt{\frac{S_{0}}{S'_{0}}} \sim - \frac{1}{2\qty(1 + \alpha)} \dv[3]{x} \ln(\func{F}{x}), \quad \qty(x \to \infty).
\end{equation}
This fantastic cancellation of the lower order terms is due to the ansatz \eqref{eq:reg1-ansatz}. On the other hand,
\begin{equation}
    \frac{\func{S'_{0}}{x}}{\func{S_{0}}{x}} \sim -\qty(1 + \alpha) + \frac{\func{F'}{x}}{\func{F}{x}}, \quad \qty(x \to \infty).
\end{equation}
Therefore, if \(\func{P}{x}\) is not asymptotic to a constant, we should choose \(\nu = 0\). As stated previously \(\func{P}{x}\) should only contain constant terms at large \(x\) if the cross section is decaying exponentially fast due to a low lying resonance or threshold. If this is the case then we should choose
\begin{equation} \label{eq:nu}
    \nu = \pm \frac{\alpha}{2 \qty(1 + \alpha)},
\end{equation}
so that the constant term cancels. We can therefore guarantee that in the worst case scenario
\begin{equation}
    \func{S_{2}}{x} \sim - \frac{1}{2} \frac{D}{x \func{Q}{x}}, \quad \qty(x \to \infty).
\end{equation}
for some constant $D$. This indicates an improvement over standard WKB, because \(\lambda \func{S_{0}}{x} = 1\) occurs when \(x = \order{\ln(\lambda)}\). At this same point the correction \(\func{S_{2}}{x} / \lambda = \order{1 / \ln(\lambda)}\) at most, and therefore our approximation extends into the region where \(\lambda \func{S_{0}}{x} \ll 1\) (FIG.~\ref{fig:S0S2}). We then define the upper bound of the thermal-equilibrium region by where the leading order term is equal in magnitude to the correction term,
\begin{equation}
    \lambda^{2} \func[2]{Q}{x_{+}} \coloneqq \frac{1}{x_{+}}.
\end{equation}
The more common scenario is \(\alpha = 0\) and \(\beta = -2\), which yields a much larger upper bound
\begin{equation}
    \lambda^{2} \func[2]{Q}{x_{+}} \coloneqq \frac{1}{x^{3}_{+}}.
\end{equation}
In any case \(x_{+} = \order{\ln(\lambda)}\) due to the exponential decay of the equilibrium abundance, so that as \(\lambda \to \infty\) the upper bound of the region of validity also goes to infinity as expected.

Finally, we have the approximation of the solution in the thermal equilibrium region:
\begin{multline} \label{eq:uI}
    \func{\uI}{x} \sim \sqrt{\frac{\int^{\infty}_{x} \dd{s} \func{Q}{s}}{\func{Q}{x}}} \func{K_{\nu}}{\lambda \int^{\infty}_{x} \dd{s} \func{Q}{s}},\\[2mm]
    \qty(\lambda \to \infty), \quad x \ll x_{+}\\[3mm]
    \nu = \begin{cases} 0 & \text{w/o res/thresh} \\[2mm] \frac{\alpha}{2 \qty(1 + \alpha)} & \text{w/ res/thresh} \end{cases}. 
\end{multline}

\subsection{Post freeze-out region (Region III)} \label{subsec:regionIII}

In the post-freeze out region, approximate \eqref{eq:udiffeq} by taking the limit \(x \to \infty\) while holding \(\lambda\) constant. Neglecting the first term at leading order in \eqref{eq:udiffeq} (which is exponentially suppressed as \(x\to\infty\)), the differential equation becomes
\begin{equation} \label{eq:uIIIdiffeq}
u'' \sim \func{P}{x} u\,,\quad \qty(x\to\infty)\\
\end{equation}
Recalling the definition of \(\func{P}{x}\) \eqref{eq:P}, we see that \eqref{eq:uIIIdiffeq} can be solved exactly, yielding
\begin{equation} \label{eq:uIIILO}
\func{\uIII}{x} \sim \frac{1}{\sqrt{\func{f}{x}}} \qty[c_{1} - c_{2} \int^{\infty}_{x} \func{f}{s} \dd{s}], \; \qty(x \to \infty).
\end{equation}
The arbitrary integration constants \(c_{1}\) and \(c_{2}\) cannot be determined because the boundary condition at \(x=0\) lies outside the region of validity of this approximation.

To obtain the higher order corrections to \eqref{eq:uIIILO} we construct a series solution of the form
\begin{equation} \label{eq:uIIIseries}
    \func{\uIII}{x} \sim \sum^{\infty}_{k = 0} \func{\uIII_{k}}{x} e^{- 2 \qty(1 + \alpha) k x}, \quad \qty(x \to \infty).
\end{equation}
The solution for \(\func{\uIII_{1}}{x}\) can be obtained directly (and in closed form) using the method of variation of parameters. For the sake of clarity we retain only the largest component:
\begin{multline}
    \func{\uIII_{1}}{x} \sim \frac{\lambda^{2} \func[2]{F}{x}}{4 \qty(1 + \alpha) \sqrt{\func{f}{x}}} \qty[c_{1} - c_{2} \int^{\infty}_{x} \func{f}{s} \dd{s}],\\[2mm] \qty(x \to \infty).
\end{multline}
As was the case in the thermal equilibrium region, we require the higher order corrections to be negligible compared to the leading order result in order to claim a valid asymptotic approximation. This requirement then defines an estimate of the lower bound on the region of validity of the post-freeze-out approximation. We again obtain a transcendental definition of the lower bound estimate \(x_{-}\):
\begin{equation}
    \lambda \func{Q}{x_{-}} = 1.
\end{equation}
Like the estimate of the upper bound of region I, \(x_{-}\) is \(\order{\ln(\lambda)}\).

Our final approximation of the solution in the post-freeze-out region is
\begin{multline} \label{eq:uIIIsolution}
    \func{\uIII}{x} \sim \frac{1}{\sqrt{\func{f}{x}}} \qty[c_{1} - c_{2} \int^{\infty}_{x} \func{f}{s} \dd{s}], \\[2mm] \qty(x \to \infty), \quad x \gg x_{-}.
\end{multline}
Inserting \eqref{eq:uIIIsolution} into \eqref{eq:y2u} and taking the limit \(x \to \infty\) yields the relic abundance
\begin{equation}
\begin{split}
    Y_{\infty} \equiv& \lim_{x \to \infty} \func{Y}{x},\\[2mm]
    =& \frac{c_{2}}{\lambda c_{1}}.
\end{split}
\end{equation}
In section \ref{sec:matching} we will approximate the coefficients \(c_{1}\) and \(c_{2}\).

\subsection{Freeze-out region (Region II)} \label{subsec:regionII}

Comparing \(x_{-}\) to \(x_{+}\) it is easy to see that there exists an overlap region where the thermal equilibrium and post-freeze-out approximations are both valid. Therefore, we can (and will) determine the constants \(c_{1}\) and \(c_{2}\) order by order by asymptotically matching the region I solution to the region III solution. However, it proves advantageous to construct an approximation in the overlap region in order to guide the asymptotic matching. We will define a \textit{freeze-out} temperature \(\xf \in \qty(x_{-}, x_{+})\) with which we can organize the asymptotic limits. Because this \(\xf\) is order \(\ln(\lambda)\) it is itself a large parameter if \(\lambda\) is large. We then construct a series solution in the overlap region by assuming
\begin{equation} \label{eq:regII_series}
    \func{\uII}{t} \sim \sum^{\infty}_{k = 0} \frac{\func{\uII_{k}}{t}}{\xf^{k}}, \quad \qty(\xf \to \infty),
\end{equation}
where \(t \equiv x - \xf\). The relic abundance will not depend on the precise definition of \(\xf\), but in order to obtain numerical values one must specify it explicitly. For now, we define the freeze-out temperature \(\xf\) to occur when
\begin{equation}
\lambda Q(x_{\mathrm{f}}) = \order{1}, \quad (x_{\mathrm{f}} \to \infty),
\end{equation}
so that the series representation \eqref{eq:regII_series} begins with an order 1 constant. Alternative definitions would require the leading order term to vanish in some cases (depending on the asymptotic form of the thermally averaged cross section in the overlap region), and our choice avoids this inconvenience.

Next we expand the differential equation \eqref{eq:udiffeq2} around \(\xf\) using
\begin{subequations}
\begin{align}
    \func{Q}{\xf + t} \sim& \func{Q}{\xf} e^{- \qty(1 + \alpha) t}, \quad \qty(\xf \to \infty),\\[2mm]
    \func{P}{\xf + t} \sim& \func{P}{\xf}, \quad \qty(\xf \to \infty),
\end{align}
\end{subequations}
for any finite \(t\) to yield
\begin{multline}
    \dv[2]{\func{u}{t}}{t} \sim \qty[\lambda^{2} \func[2]{Q}{\xf} e^{- 2 \qty(1 + \alpha) t} + \func{P}{\xf}] \func{u}{t},\\[2mm] \qty(\xf \to \infty).
\end{multline}
The solutions are linear combinations of modified Bessel functions. However, it is usually the case that we should not retain the \(\func{P}{\xf}\) term. If the annihilation cross section does not vanish at threshold then \(\func{P}{\xf}\) is at most of order \(\order{\xf^{-2}}\). We can enforce this distinction by allowing for two cases: \(\alpha \ll 1 / \sqrt{\xf}\) and \(\alpha \gtrsim 1 / \sqrt{\xf}\). The leading order solution is then
\begin{equation}
    \func{\uII_{0}}{t} = b_{1} \func{K_{\nu}}{\Lambda_{\mathrm{f}} e^{- \qty(1 + \alpha) t}} + b_{2} \func{I_{\nu}}{ \Lambda_{\mathrm{f}} e^{- \qty(1 + \alpha) t}},
\end{equation}
where \(\Lambda_{\mathrm{f}} \equiv \lambda \func{Q}{\xf} / \qty(1 + \alpha)\) and
\begin{equation}
    \nu = \begin{cases} 0 & \alpha \ll \frac{1}{\sqrt{\xf}} \\[2mm] \frac{\sqrt{\func{P}{\xf}}}{1 + \alpha} = \frac{\alpha}{2 \qty(1 + \alpha)} & \alpha \gtrsim \frac{1}{\sqrt{\xf}} \end{cases}
\end{equation}
Note the exact agreement of the parameter \(\nu\) as derived in section \ref{subsec:regionI}.

The leading order matching between region I and II is obvious:
\begin{subequations}
\begin{align}
    b_{1} =& \frac{1}{\sqrt{1 + \alpha}},\\[2mm]
    b_{2} =& 0.
\end{align}
\end{subequations}
Because our choice of the freeze out temperature \(\xf\) lies near the lower bound of the overlap region we must also take \(t \to \infty\). The solution in the overlap region is then approximately linear if \(\alpha \ll 1 / \sqrt{\xf}\) or a sum of exponential terms \(\exp(\alpha t / 2)\) and \(\exp(- \alpha t / 2)\) if \(\alpha \gtrsim 1 \sqrt{\xf}\). One could have chosen to define the freeze-out condition differently, and the behavior of the solution in the overlap region would be identical, but we could not have written it in such a simple way.

What we have learned is that, if we neglect \(\alpha\), the constant term and the term proportional to \(t\) must be considered the same order. Similarly, if \(\alpha\) is not neglected, the exponential terms should also be considered the same order.

\section{Asymptotic Matching} \label{sec:matching}

With asymptotic approximations in hand for the thermal-equilibrium region and post-freeze-out region we now asymptotically match the solutions in the region where both approximations are valid. We will utilize the following approximations of \(\func{S_{0}}{x}\):
\begin{widetext}
\begin{subequations}
\begin{align}
    \label{eq:S0xfseries}
    \func{S_{0}}{\xf + t} \sim& \; \func{Q}{\xf} e^{-\qty(1 + \alpha) t} \sum^{\infty}_{j = 0} \sum^{j}_{k = 0} \frac{\qty[\qty(1 + \alpha) t]^{k}}{k! \qty(1 + \alpha)^{j + 1}} \frac{1}{\func{F}{\xf}} \eval{\dv[j]{\func{F}{x}}{x}}_{x = \xf}, \quad \qty(\xf \to \infty),\\[2mm]
    \label{eq:S0primexfseries}
    \func{S'_{0}}{\xf + t} \sim& \; - \func{Q}{\xf} e^{-\qty(1 + \alpha) t} \sum^{\infty}_{j = 0} \frac{t^{j}}{j!} \frac{1}{\func{F}{\xf}} \eval{\dv[j]{\func{F}{x}}{x}}_{x = \xf}, \quad \qty(\xf \to \infty).
\end{align}
\end{subequations}
\end{widetext}
These can be found by taking the Taylor expansion of \(\dv*[j]{\func{F}{x}}{x}\) around \(\xf + t\) in \eqref{eq:S0series}. These representations are convenient because the sum over \(j\) yields a series in increasing powers of \(1/\xf\). We split the matching procedure into three categories: leading order assuming \(\alpha = 0\), next to leading order assuming \(\alpha = 0\), and leading order for general \(\alpha\). Because our choice for the freeze-out condition is near the lower bound of the overlap region there will not be a true leading order matching condition for the \(\alpha = 0\) case. What we label as leading order is in fact next to leading order, and what we have labeled as next to leading order is actually next to next to leading order.

\subsection{Leading Order} \label{subsec:matchingLO}

Assuming that \(\alpha\) is either large enough that resonance and threshold effects are negligible in the overlap region or that \(\alpha\) is of order \(1/\xf\) or smaller we shift the dependent variable by \(x \equiv \xf + t\). In region I we retain only the leading order terms in \eqref{eq:S0xfseries} and \eqref{eq:S0primexfseries}.
\begin{equation}
    \func{\uI}{t} \sim C + t, \quad \qty(\xf \to \infty),
\end{equation}
Where
\begin{equation}
    C \equiv - \ln(\frac{\lambda \func{Q}{\xf}}{2}) - \gamma,
\end{equation}
is an order 1 constant and \(\gamma\) is the Euler-Mascheroni constant. To obtain this approximation we have taken the limit \(t \to \infty\) and used the small argument expansion of the modified Bessel function (with \(nu = 0\)). In addition, there are terms that are exponentially suppressed at large \(t\), but these can be neglected at leading order. Similarly, in region III we have
\begin{multline} \label{eq:uIIILOregII}
    \func{\uIII}{t} \sim \frac{1}{\sqrt{\func{f}{\xf}}} \qty[c_{1} - c_{2} \int^{\infty}_{\xf} \func{f}{s} \dd{s} + c_{2} \func{f}{\xf} t],\\[2mm] \qty(\xf \to \infty).
\end{multline}
It may seem odd that the linear term in \(t\) is retained, because it is down by one power of \(\xf\) compared to the second constant term. However, as we learned in Section \ref{subsec:regionII}, the constant term and the term proportional to \(t\) must be considered the same order. Large terms will cancel between the \(c_{1}\) term and \(c_{2}\) term, so that overall the constant term is of the same order as the term linear in \(t\). It is then simple to determine the constants \(c_{1}\) and \(c_{2}\).
\begin{subequations}
\begin{align}
    c_{1} \sim& \frac{1}{\sqrt{\func{f}{\xf}}} \qty[\int^{\infty}_{\xf} \func{f}{s} \dd{s} + C \func{f}{\xf}], \; \qty(\xf \to \infty),\\[2mm]
    c_{2} \sim& \frac{1}{\sqrt{\func{f}{\xf}}}, \quad \qty(\xf \to \infty).
\end{align}
\end{subequations}
Inserting these approximations into our expression for the relic abundance yields our leading order approximation:
\begin{equation} \label{eq:YinfinityLO}
    Y_{\infty} \sim \frac{1}{\lambda \int^{\infty}_{\xf} \func{f}{s} \dd{s} + C \lambda \func{f}{\xf}}, \quad \qty(\xf \to \infty).
\end{equation}

So far we have derived the leading order asymptotic approximation of the relic abundance without specifying an exact value for \(\xf\). In fact, these results do not depend strongly on the precise value of \(\xf\). Allow \(\xf \to \xf + \varepsilon\), where \(\varepsilon \ll \xf\). Under this shift
\begin{subequations}
\begin{align}
    \func{Q}{\xf} \to& \, \func{Q}{\xf} e^{- \varepsilon} \qty(1 + \order{\frac{\varepsilon}{\xf}})\\[2mm]
    C \to& \: C + \varepsilon + \order{\frac{\varepsilon}{\xf}}
\end{align}
\end{subequations}
The ratio of the region III coefficients then transforms as
\begin{equation}
\begin{split}
    \frac{c_{1}}{c_{2}} \to& \int^{\infty}_{\xf + \varepsilon} \func{f}{s} \dd{s} + \qty(C + \varepsilon) \func{f}{\xf}\\[2mm]
    =& \int^{\infty}_{\xf} \func{f}{s} \dd{s} + C \func{f}{\xf},
\end{split}
\end{equation}
which shows that the relic abundance is invariant under a small shift of the freeze-out temperature up to \(\order{\varepsilon^{2} / \xf^{2}}\).

\subsection{Next to Leading Order} \label{subsec:matchingNLO}

At next to leading order we retain terms up to \(1/\xf\) and \(t / \xf\), but continue to drop terms like \(1/\xf^{2}\) and \(\exp(-t)\). The approximations in each region become:
\begin{equation}
    \func{\uI}{t} \sim \alpha_{1} + \beta_{1} t, \quad \qty(\xf \to \infty).
\end{equation}
\begin{equation}
    \func{\uIII}{t} \sim \alpha_{3} + \beta_{3} t, \quad \qty(\xf \to \infty).
\end{equation}
Where the coefficients are 
\begin{subequations}
\begin{align}
    \alpha_{1} \equiv& \qty[1 + \frac{1}{2} \frac{\func{F'}{\xf}}{\func{F}{\xf}} ] C - \frac{\func{F'}{\xf}}{\func{F}{\xf}},\\[2mm]
    \beta_{1} \equiv& 1 - \frac{1}{2} \frac{\func{F'}{\xf}}{\func{F}{\xf}},
\end{align}
\begin{align}
    \alpha_{3} \equiv& \frac{1}{\sqrt{\func{f}{\xf}}} \qty[ c_{1} - c_{2} \int^{\infty}_{\xf} \func{f}{s} \dd{s}],\\[2mm]
    \beta_{3} \equiv& c_{2} \sqrt{\func{f}{\xf}} - \frac{1}{2} \frac{\func{f'}{\xf}}{\func{f}{\xf}} \alpha_{3}
\end{align}
\end{subequations}
Note the lack of a \(t^{2}\) term in the region III solution, it has cancelled exactly. After a little algebra one can simultaneously solve for the coefficients \(c_{1}\) and \(c_{2}\).
\begin{widetext}
\begin{subequations}
\begin{align}
    c_{1} \sim& \frac{1}{\sqrt{\func{f}{\xf}}} \qty{\qty[1 + \frac{C}{2} \frac{\func{f'}{\xf}}{\func{f}{\xf}} - \frac{1}{2} \frac{\func{F'}{\xf}}{\func{F}{\xf}}] \int^{\infty}_{\xf} \func{f}{s} \dd{s} + \func{f}{\xf} \qty[C + \frac{1}{2} \frac{\func{F'}{\xf}}{\func{F}{\xf}} \qty(C - 2)]},\\[2mm]
    c_{2} \sim& \frac{1}{\sqrt{\func{f}{\xf}}} \qty[1 + \frac{C}{2} \frac{\func{f'}{\xf}}{\func{f}{\xf}} - \frac{1}{2} \frac{\func{F'}{\xf}}{\func{F}{\xf}}].
\end{align}
\end{subequations}
\begin{equation} \label{eq:YinfinityNLO}
    Y_{\infty} \sim \frac{1}{\lambda} \qty{\int^{\infty}_{\xf} \func{f}{s} \dd{s} + C \func{f}{\xf} - \frac{C^{2}}{2} \func{f'}{\xf} + \qty(C - 1) \func{f}{\xf} \frac{\func{F'}{\xf}}{\func{F}{\xf}}}^{-1}, \quad \qty(\xf \to \infty).
\end{equation}
\end{widetext}
We again check to ensure that the relic abundance does not depend strongly on the exact choice of freeze-out temperature. Shifting \(\xf \to \xf + \varepsilon\), retaining the \(\order{\epsilon^{2} / \xf^{2}}\) term, and using
\begin{equation}
    C \to C + \epsilon - \epsilon \frac{\func{F'}{\xf}}{\func{F}{\xf}} + \order{\frac{\varepsilon^{2}}{\xf^{2}}},
\end{equation}
we find that the \(\order{\epsilon / \xf}\), \(\order{\epsilon / \xf^{2}}\), and \(\order{\epsilon^{2} / \xf^{2}}\) all cancel identically in the relic abundance. Therefore, we make the convenient choice for the freeze-out temperature of \(C = 1\). This choice defines the numerical value of the freeze-out temperature by
\begin{equation} \label{eq:freezeout}
    \lambda \func{Q}{\xf} = 2 e^{- \gamma - 1}.
\end{equation}
The third term in \eqref{eq:YinfinityNLO} then vanishes identically, and the remaining three terms match exactly to
\begin{multline}
    \int^{\infty}_{\xf - 1} \func{f}{s}{\dd{s}} \sim \int^{\infty}_{\xf} \func{f}{s} \dd{s} + \func{f}{\xf} - \frac{1}{2} \func{f'}{\xf},\\[2mm] \qty(\xf \to \infty).
\end{multline}
It is then a straightforward numerical exercise to determine the relic abundance up to order \(1/\xf^{3}\). One simply determines the freeze-out temperature using \eqref{eq:freezeout} and then integrates the thermally averaged cross section (with the appropriate cosmological factors) from \(\xf - 1\) to infinity.

This result is very similar to those in the literature, with some seemingly minor but important corrections. Writing
\begin{equation}
    \func{f}{x} \sim x^{-n -2} \sum^{\infty}_{k = 0} f_{k} x^{-k}, \: f_{0} = 1, \quad \qty(x \to \infty),
\end{equation}
the relic abundance can be written
\begin{equation} \label{eq:Yinf_asymp_f}
    Y_{\infty} \sim \frac{\qty(n+1) \xf^{n+1}}{\lambda \qty[1 + \frac{\Upsilon_{1}}{\xf} + \frac{\Upsilon_{2}}{\xf^{2}}]}, \quad \qty(\xf \to \infty),
\end{equation}
with
\begin{subequations}
\begin{align}
    \Upsilon_{1} \equiv& \frac{\qty(n+1) \qty(n+2) + f_{1}}{\qty(n+2)}\\[2mm]
    \Upsilon_{2} \equiv& \frac{\qty(n+1) \qty(n + 2) \qty(n+3) + 2 \qty(n + 3) f_{1} + 2 f_{2}}{2 \qty(n + 3)}
\end{align}
\end{subequations}
Dropping all but the first term in the denominator yields a result of the same form as in \cite{kolb1994early}, but with a different choice for the freeze-out temperature. However, the \(1/\xf\) term is what guarantees that the result does not depend strongly on the choice of freeze-out temperature. The error then depends linearly on the choice of \(\xf\), which indicates that the approximation is, strictly speaking, invalid.

Keeping the \(1/\xf\) corrections in \eqref{eq:Yinf_asymp_f} reproduces the results of \cite{Bender2012} after correcting for mistakes in their analysis. This gives us confidence that boundary-layer-analysis can be used to construct approximate solutions to other Boltzmann equations.


\subsection{Including Resonance and Threshold Effects} \label{subsec:matchingResThresh}

We next assume that the thermally averaged cross section is exponentially decaying at leading order, with the coefficient in the exponent, \(\alpha\), being much larger than \(1/\xf\). In order to accommodate the additional Boltzmann suppression we write
\begin{equation}
    \func{f}{x} \sim \func{g}{x} e^{- \alpha x}, \quad \qty(x \to \infty).
\end{equation}
Much like \(\func{F}{x}\) we assume we have factored out all the exponential behavior so that \(\func{g}{x}\) has a valid asymptotic approximation in powers of \(1/x\) as \(x \to \infty\). We will again shift the dependent variable to \(x = \xf + t\), and to further approximate the region III solution we split the integral into two parts,
\begin{equation}
    \int^{\infty}_{\xf + t} \func{f}{s} \dd{s} \sim \int^{\infty}_{\xf} \func{f}{s} \dd{s} - \int^{\xf + t}_{\xf} \func{f}{s} \dd{s}.
\end{equation}
It is necessary to split the integral because in general the thermally averaged cross section will not be well approximated by this exponential behavior if \(x\) is sufficiently large for any finite set of parameters. We therefore leave the first integral to be evaluated numerically. The second integral can be evaluated to all orders assuming \(\func{g}{x}\) is a slowing varying function over the range of integration.
\begin{equation} \label{eq:intf_series}
    \int^{\xf + t}_{\xf} \func{f}{s} \dd{s} \sim \sum^{\infty}_{k = 0} \frac{e^{- \alpha \xf}}{\alpha^{k + 1}} \func{\gamma}{k + 1, \alpha t} \dfunc[n]{g}{\xf},
\end{equation}
where \(\dfunc[n]{g}{\xf} \equiv \eval{\dv*{\func{g}{x}}{x}}_{x = \xf}\) and \(\func{\gamma}{x, z}\) is the lower incomplete gamma function,
\begin{equation}
    \func{\gamma}{s, z} \coloneqq \int^{z}_{0} x^{s - 1} e^{-x} \dd{x}.
\end{equation}
Each term in the series is suppressed by \(1/ \qty(\alpha \xf)\) if \(\alpha\) is large. If \(\alpha\) is small the incomplete gamma function goes like \(\qty(\alpha t)^{k + 1}\), which cancels all the factors of \(\alpha\) in the denominator. In either case we can further approximate the region III solution by retaining only the first term in the series \eqref{eq:intf_series}:
\begin{equation}
    \func{\uIII}{\xf + t} \sim \frac{1}{\sqrt{\func{f}{\xf}}} \qty[\func{\theta^{-}_{\alpha}}{\xf} e^{\frac{\alpha t}{2}} - \func{\theta^{+}_{\alpha}}{\xf} e^{- \frac{\alpha t}{2}}]
\end{equation}
where the constants are,
\begin{subequations}
\begin{align}
    \func{\theta^{-}_{\alpha}}{\xf} \equiv& c_{1} - c_{2} \int^{\infty}_{\xf} \func{f}{s} \dd{s} + \frac{c_{2}}{\alpha} \func{f}{\xf},\\[2mm]
    \func{\theta^{+}_{\alpha}}{\xf} \equiv& \frac{c_{2}}{\alpha \sqrt{\func{f}{\xf}}}.
\end{align}
\end{subequations}

Similarly, the region I approximation becomes
\begin{multline}
    \func{\uI}{t} \sim \frac{1}{\sqrt{1 + \alpha}} \func{K_{\nu}}{\frac{\lambda \func{Q}{\xf} e^{- \qty(1 + \alpha) t}}{1 + \alpha}},\\[2mm] \qty(\xf \to \infty),
\end{multline}
where \(\nu\) is defined by \eqref{eq:nu}. We note that, for any value of \(\alpha\), the order of the Bessel function \(\nu \in \qty[0,\tfrac{1}{2}]\), we therefore let \(t \to \infty\) and use the small argument expansion of the Bessel function for non integral orders.
\begin{multline}
    \func{\uI}{t} \sim B_{\alpha} \qty[\func{\eta^{-}_{\alpha}}{\xf} e^{\alpha t / 2} - \func{\eta^{+}_{\alpha}}{\xf} e^{- \alpha t / 2}], \\[2mm] \qty(\xf \to \infty),
\end{multline}
where the constants are
\begin{subequations}
\begin{align}
    \func{B_{\alpha}}{\xf} \equiv& \frac{\pi}{2 \sqrt{1 + \alpha} \sin(\frac{\alpha \pi}{2 \qty(1 + \alpha)})}\\[2mm]
    \func{\eta^{+}_{\alpha}}{\xf} \equiv& \frac{1}{\func{\Gamma}{\frac{2 + 3 \alpha}{2 + 2 \alpha}}} \qty(\frac{\lambda \func{Q}{\xf}}{2 \qty(1 + \alpha)})^{\frac{\alpha}{2 \qty(1 + \alpha)}}\\[2mm]
    \func{\eta^{-}_{\alpha}}{\xf} \equiv& \frac{1}{\func{\Gamma}{\frac{2 + \alpha}{2 + 2 \alpha}}} \qty(\frac{\lambda \func{Q}{\xf}}{2 \qty(1 + \alpha)})^{- \frac{\alpha}{2 \qty(1 + \alpha)}} 
\end{align}
\end{subequations}
Both solutions exhibit the exact same exponential behavior. The coefficients \(c_{1}\) and \(c_{2}\) are easily found:
\begin{subequations}
\begin{align}
    c_{1} \sim& c_{2} \qty{\int^{\infty}_{\xf} \func{f}{s} \dd{s} - \frac{\func{f}{\xf}}{\alpha} \qty[1 - \frac{\func{\eta^{-}}{\xf}}{\func{\eta^{+}}{\xf}}]},\\[2mm]
    c_{2} \sim& \frac{\alpha B_{\alpha} \func{\eta^{+}_{\alpha}}{\xf}}{\sqrt{\func{f}{\xf}}}, \quad \qty(\xf \to \infty).
\end{align}
\end{subequations}

To simplify the notation and computational determination of the relic abundance we next define the parameter
\begin{equation}
    \delta x \coloneqq \frac{1}{\alpha} \ln(\frac{\func{\eta^{-}_{\alpha}}{\xf}}{\func{\eta^{+}_{\alpha}}{\xf}}).
\end{equation}
This parameter has the following asymptotic behavior:
\begin{subequations}
\begin{align}
    \delta x \sim& 1 + \tfrac{1}{24} \qty(\dfunc[2]{\psi}{1} - 12) \alpha^{2}, \quad \qty(\alpha \to 0),\\[2mm]
    \delta x \sim& \frac{1 + \gamma - \ln(2) + \ln(\alpha)}{\alpha}, \quad \qty(\alpha \to \infty),
\end{align}
\end{subequations}
where \(\dfunc[m]{\psi}{z}\) is the polygamma function of order \(m\). Using this parameter we may write the relic abundance as
\begin{equation} \label{eq:relic_abun_gen}
    Y_{\infty} \sim \qty[\lambda \int^{\infty}_{\xf - \delta x} \func{f}{s} \dd{s}]^{-1}, \quad \qty(\xf \to \infty).
\end{equation}

The result \eqref{eq:relic_abun_gen} is valid for all values of \(\alpha\), and in the limit \(\alpha \to 0\) reproduces the results of the previous section. It is correct up to \(1/\xf^{2}\) corrections for general \(\alpha\) and up to \(1/\xf^{3}\) corrections when \(\alpha \ll 1 / \xf\).

\section{Results} \label{sec:results}

We have determined an asymptotic approximation of the relic abundance in the limit that the number density of the particle species is very nearly its thermal equilibrium value until \(T \ll m\), where \(m\) is the mass of the particle. We define the freeze-out condition as
\begin{equation} \label{eq:results:freeze-out}
    \sqrt{\frac{\pi}{45}} \frac{m \mplanck \func[1/2]{g_{*, \mathrm{eff}}}{\xf}}{\xf^{2}} \expval{\sigma \vmoller} \func{\yeq}{\xf} = 2 e^{- 1 - \gamma}.
\end{equation}
The asymptotic approximation of the relic abundance is
\begin{multline} \label{eq:results:Yinf}
Y_{\infty} \sim \frac{\sqrt{45}}{\sqrt{\pi} m \mplanck} \qty[\int^{\infty}_{\xf - \delta x} \frac{\func[1/2]{g_{*, \mathrm{eff}}}{s}}{s^{2}} \expval{\sigma \vmoller} \dd{s}]^{-1}\\[2mm]\qty(\xf \to \infty),
\end{multline}
where the shift in the integration range \(\delta x\) is given by
\begin{equation} \label{eq:results:deltax}
    \delta x = \frac{1}{\alpha} \ln(\frac{\func{\Gamma}{\frac{2 + 3 \alpha}{2 + 2 \alpha}}}{\func{\Gamma}{\frac{2 + \alpha}{2 + 2 \alpha}}}) + \frac{1 + \gamma + \ln(1 + \alpha)}{1 + \alpha}.
\end{equation}
In order to apply this approximation one must have some knowledge of the analytic behavior of the thermally averaged annihilation cross section in the vicinity of \(\xf\). If, as is usually the case, the thermally averaged cross section behaves like some power of \(1/x\) near \(\xf\) then one should set \(\alpha=0\), i.e. \(\delta x = 1\). On the other hand, if the leading order behavior near \(\xf\) of the annihilation cross section has an exponential character due to resonance or threshold effects, i.e.
\begin{equation*}
    \expval{\sigma \vmoller} \sim x^{\beta} e^{- \alpha x},
\end{equation*}
then one should use the coefficient in the exponent, \(\alpha\), to determine \(\delta x\) from \eqref{eq:results:deltax}. We have provided the most common expressions for \(\alpha\) in \eqref{eq:alpha}.

In order to estimate the fitness of our results we next compare our approximation to a numerical determination of the relic density using a benchmark model, which we now outline.

\subsection{Benchmark Model}
The benchmark model we will use is a simple extension of the SM in which we add a massive vector boson which kinetically mixes with the SM photon and a DM fermion. The Lagrangian is given by:
\begin{subequations}
\begin{align}
    \mathcal{L} &= \mathcal{L}_{\mathrm{SM}}  +\mathcal{L}_{\mathrm{kin}} + \mathcal{L}_{\mathrm{int}}\\
    \mathcal{L}_{\mathrm{kin}} &= -\tfrac{1}{4}V_{\mu\nu}V^{\mu\nu} + \tfrac{1}{2}M_{V}V_{\mu}V^{\mu} +\overline{\chi}\left(i\cancel{\partial}-m_{\chi}\right)\chi\\
    \mathcal{L}_{\mathrm{int}} &= \dfrac{\epsilon}{2}B_{\mu\nu}V^{\mu\nu} + g V_{\mu}\overline{\chi}\gamma^{\mu}\chi
\end{align}
\end{subequations}
where $V_{\mu}$ is the new massive vector boson (with mass $M_{V}$), $\chi$ is the DM Dirac fermion (with mass $m_{\chi}$) and $B_{\mu}$ is the hyper-charge gauge boson. We take the $\chi-V$ coupling $g$ to be $\order{1}$ and the kinetic mixing parameter $\epsilon \ll 1$. The \(V\)-\(B\) mass matrix can be diagonalized by shifting $B_{\mu}\to B_{\mu} + \epsilon V_{\mu}$ and neglecting terms of $\order{\epsilon^2}$. After shifting the hyper-charge gauge boson, the vector mediator obtains interactions with the hyper-charge current:
\begin{align}
    \mathcal{L}_{\mathrm{int}} &\supset \epsilon g' J^{\mu}_{Y}V_{\mu}\notag\\
    &= g' \epsilon V_{\mu}\left(\sum_{i}Q_{i}\overline{\psi}_{i}\gamma^{\mu}\psi + \sum_{i}\overline{\psi}^{L}_{i}\gamma^{\mu}T_{3}\psi^{L}_{i}\right)
\end{align}
where the first sum runs over all SM fermions $\psi_{i}$, the second over left-handed fermions $\psi^{L}_{i}$, and \(T_{3}\) is essentially the third Pauli matrix $T_{3}=\sigma_{3}/2$. 

\begin{figure*}
    \begin{minipage}[t]{0.3\textwidth}
        \centering
        \includegraphics[width=0.8\textwidth]{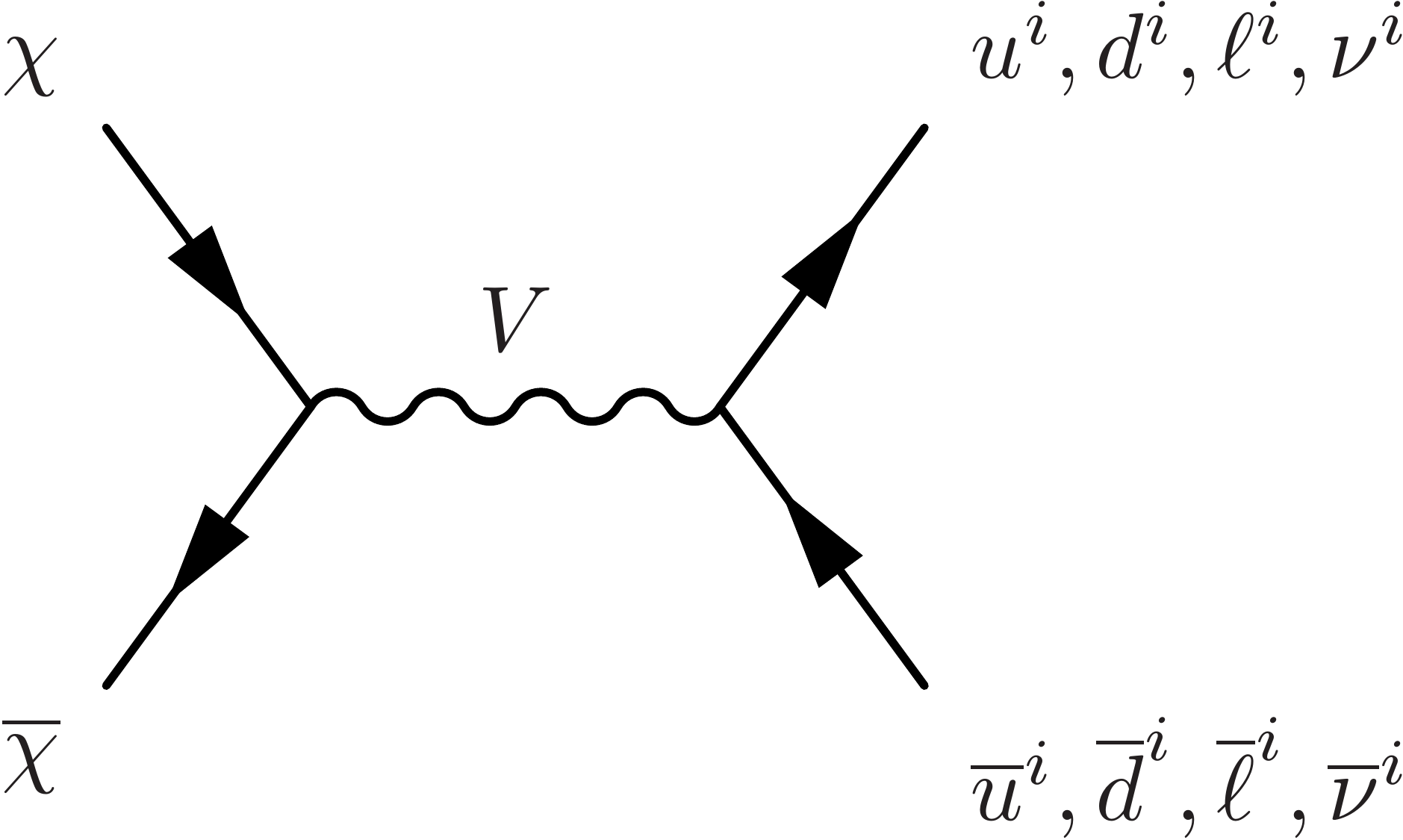}
        \label{fig:dm_ann_ffbar}
    \end{minipage}%
    \begin{minipage}[t]{0.3\textwidth}
        \centering
        \includegraphics[width=0.65\textwidth]{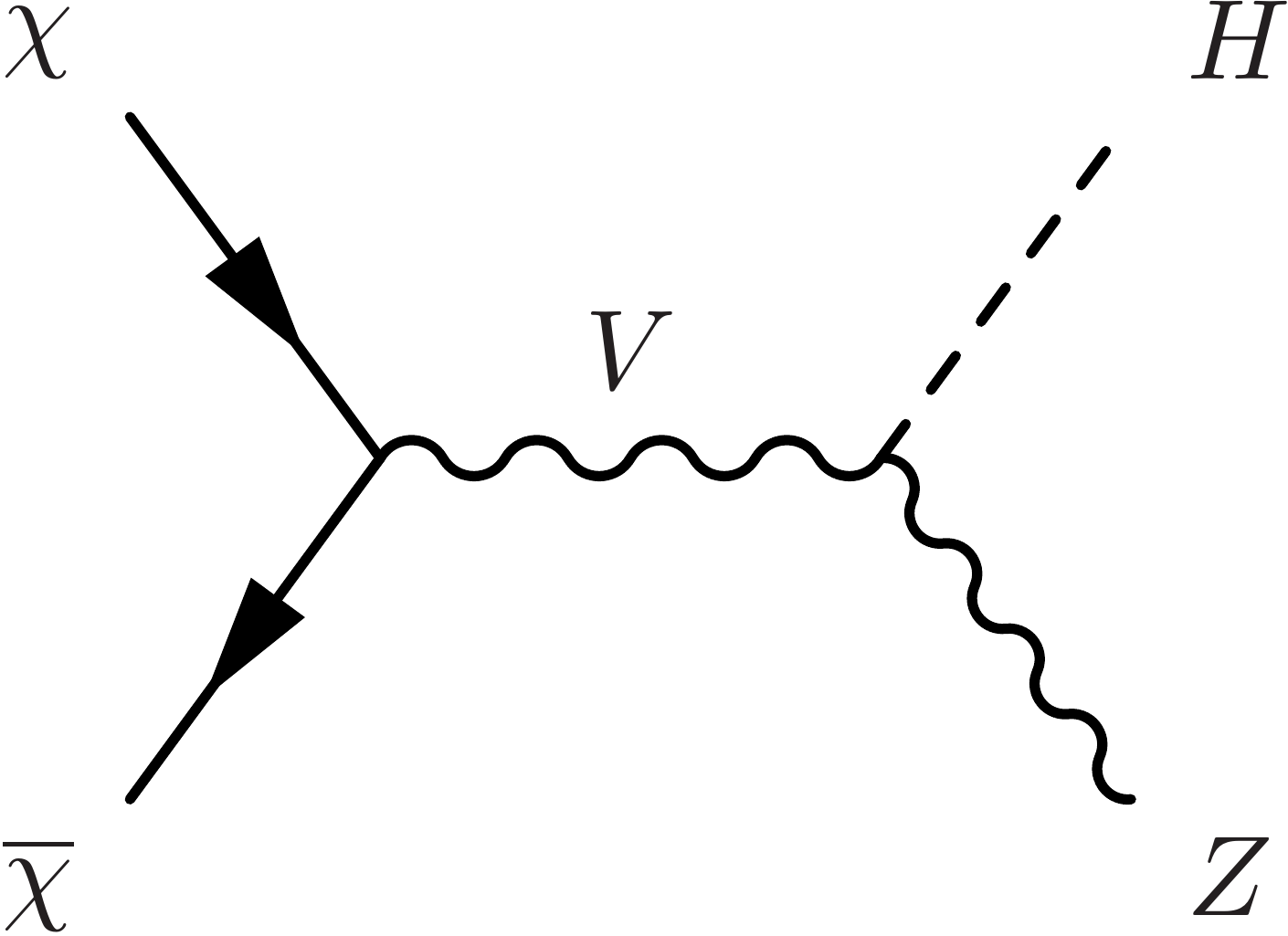}
        \label{fig::dm_ann_hz}
    \end{minipage}%
    \begin{minipage}[t]{0.25\textwidth}
        \centering
        \includegraphics[width=0.75\textwidth]{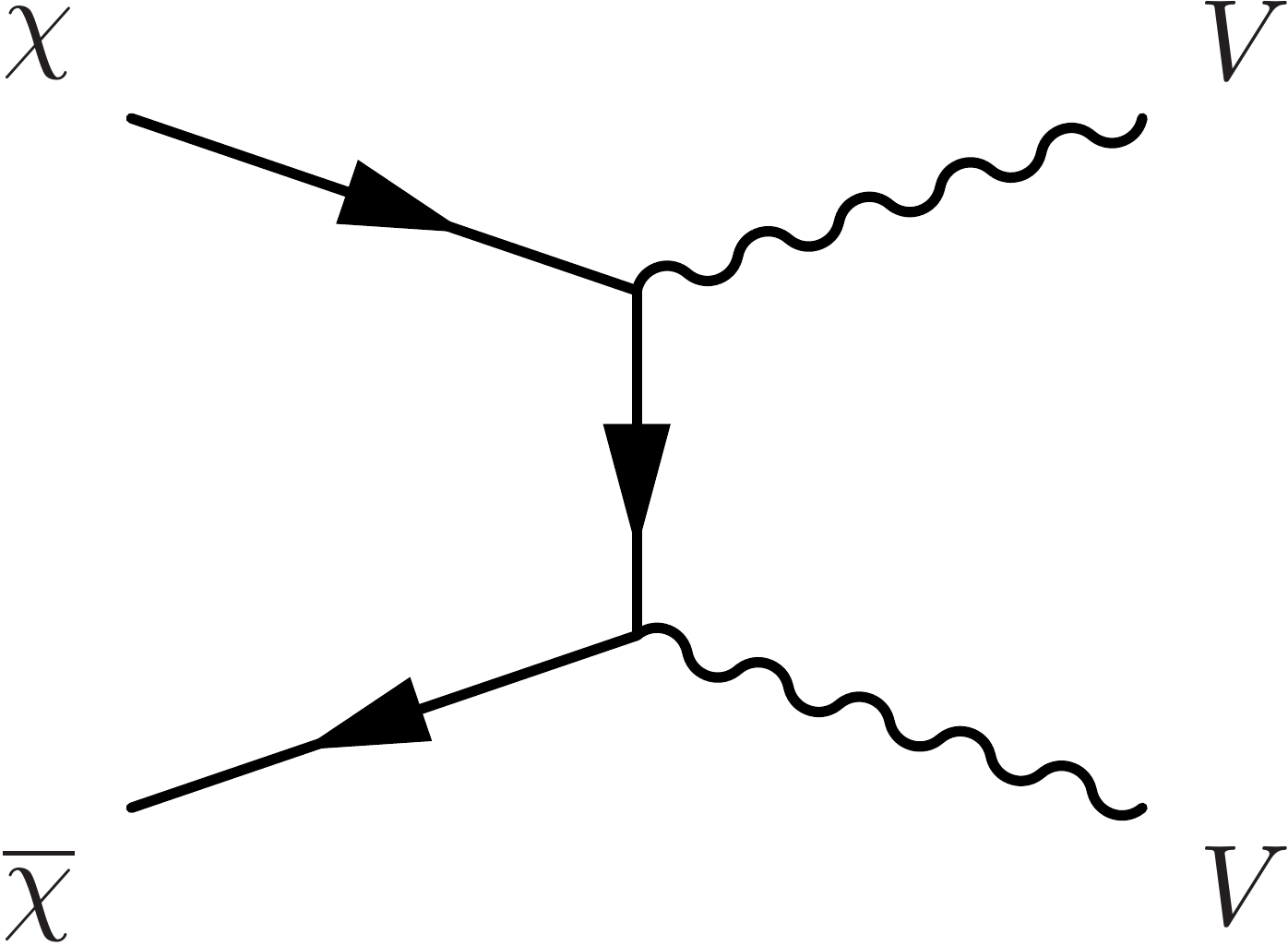}
        \label{fig:dm_ann_vv}
    \end{minipage}
\caption{Feynman diagrams showing all possible $2\to2$ annihilation channels for DM in the benchmark model.}\label{fig:dm_ann}
\end{figure*}

The thermally averaged \(2 \to 2\) annihilation cross section for \(\bar{\chi} \chi \to \text{any}\) for large \(x\) is given by
\begin{equation}
    \expval{\sigma \vmoller} \sim \int^{\infty}_{2} \dd{z} \func{\mathscr{K}}{x, z} \sum_{\vb{X}} \func{\sigma_{\bar{\chi} \chi \to \vb{X}}}{m_{\chi} z},
\end{equation}
where the thermal kernel \(\func{\mathscr{K}}{x, z}\) is
\begin{equation}
    \func{\mathscr{K}}{x, z} \equiv \frac{x}{4 \func[2]{K_{2}}{x}} z^{2} \qty(z^{2} - 4) \func{K_{1}}{x z}.
\end{equation}
In the above expressions, $z$ is the center-of-mass energy divided by the DM mass ($z\equiv\sqrt{s}/m_{\chi}$). In Fig.~\ref{fig:dm_ann} we give all possible final states.

There are three interesting regions in parameters space for this model:
\begin{enumerate}
    \item $m_{\chi} > M_{V}$: The DM is heavier than the vector mediator. There are neither thresholds nor any resonances. The dominant process is simply $\bar{\chi}\chi\to V V$. All other processes are negligible (assuming $\epsilon$ is small). \label{enum:above_thresh}
    \item $M_{V}/2 < m_{\chi} < M_{V}$: The DM is lighter than the vector but heavier than half the vector mass. At large temperatures we will pass through a threshold in which, due to finite temperature, the final state $\bar{\chi}\chi\to V V$ opens up. For smaller temperatures, this final state becomes Boltzmann suppressed. \label{enum:below_thesh_above_res}
    \item $m_{\chi} < M_{V}/2$: The DM is lighter than half the vector mass. At large temperatures, we will pass through both a resonance ($z=m_{V}/m_{\chi}$) and a threshold ($z = 2m_{V}/m_{\chi}$). \label{enum:below_res}
\end{enumerate}

\begin{figure}[ht!]
\includegraphics[width=\columnwidth]{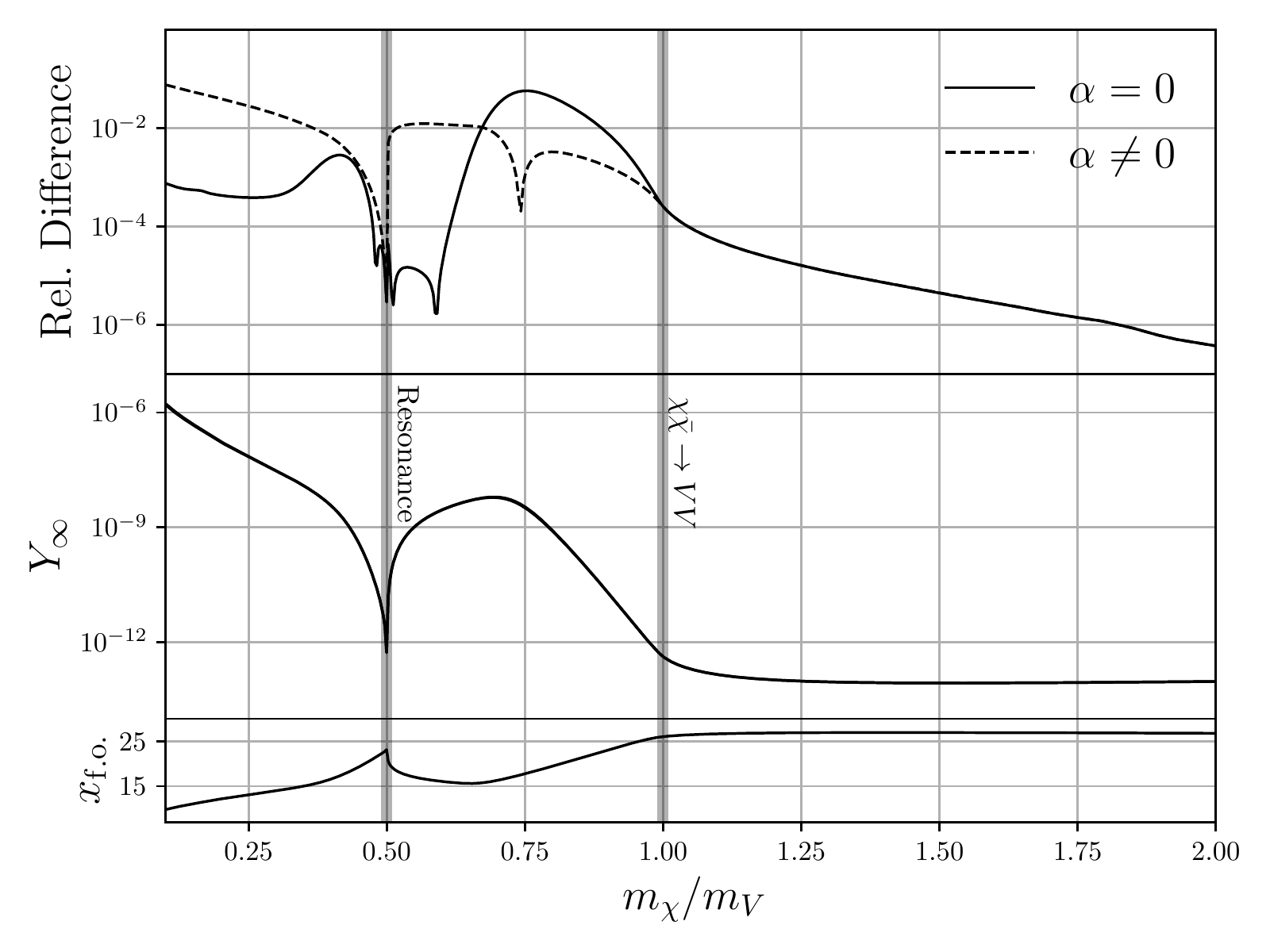}
\caption{The asymptotic approximation of the thermal relic density compared to numerical results using a benchmark model. Top: The magnitude of the relative error between the asymptotic approximation  and numerical results. Middle: The thermal relic density. Bottom: the freeze-out temperature. We set \(M_{V} = 1 \, \text{TeV}\), \(g = 1\), \(\epsilon = 10^{-3}\), and vary the mass of the DM particle from \(m_{\chi} = 100 \text{GeV} - 2 \, \text{TeV}\). One can see the affect of the resonance and \(\chi \bar{\chi} \to V V\) threshold from the rise in the power law (\(\alpha = 0\)) approximation near \(m_{\chi} / M_{V} = 0.5\) and \(m_{\chi} / M_{V} = 1\).\label{fig:relicdensity}}
\end{figure}

In Fig.~\ref{fig:relicdensity} we show the magnitude of the relative error between a numerical determination of the thermal relic density and the asymptotic approximations derived here using
\begin{equation}
    \Omega_{\chi} h^{2} = 2.744 \times 10^{8} \frac{m_{\chi}}{\text{GeV}} Y_{\infty}.
\end{equation}
The numerical results were obtained using the high-fidelity, order-switching, implicit RADAU integrator~\cite{hairer1999stiff} taken from the author's website\footnote{We use a slightly modified version of the C++ code from: \url{https://unige.ch/~hairer/software.html}.}. We recast the Boltzmann equation into a logarithmic form in order to work with numbers of $\order{1-10}$:
\begin{multline}
    \dv{W}{\ln(x)} = -\sqrt{\dfrac{\pi}{45}}\dfrac{m M_{\mathrm{pl}}}{x}g^{1/2}_{*,\mathrm{eff}}\expval{\sigma \vmoller}\\[2mm]
    \times \qty(e^{W}-e^{2W_{\mathrm{eq}}-W}),
\end{multline}
with $W = \ln(Y)$ (and $W_{\mathrm{eq}} = \ln(Y_{\mathrm{eq}})$). The integration was performed over the interval $x=1$ to $x=5 \times 10^4$, beginning the integration with $W(x=1) = W_{\mathrm{eq}}(x=1)$ and maintaining a local error of $\order{10^{-10}}$ (and a global error $\order{10^{-9}}$). In order to reduce roundoff error we employ long double (80 bit floating point) precision arithmetic.

We vary the DM mass while keeping all other parameters fixed. Because resonance effects may be important for some values of the mass ratio we compare results using \(\alpha = 0\) for all masses and the value of \(\alpha\) obtained from \eqref{eq:alpha}. With the exception of resonance and theshold effects, not accounted for in the \(\alpha = 0\) approximation, as \(\xf\) becomes larger the relative error decreases, as is expected from the asymptotic nature of the approximation.



\section{Conclusion}

We have shown, using this benchmark model, that our results satisfy the requirements of an asymptotic approximation. The controlling parameter is \(\xf\), and as \(\xf\) becomes large the relative error approaches 0. As well, our approximation yields outstanding results, giving sub percent relative errors for all parameters investigated. This is comparable or greatly exceeds the current measurement uncertainty of the Hubble parameter of roughly a percent or more~\cite{Megamaser2020,Strides2019,Sharp_Holicow_2019,Dutta_2019,Reid_2019,Dom_nguez_2019,collaboration2018planck}. The asymptotic approximation of the thermal relic density typically takes orders of magnitudes less time to compute than numerically integrating \eqref{eq:lee-weinberg}, making scans over models with large numbers of parameters more feasible. For the choices of parameters shown we typically have \(\lambda \approx 10^{14}\), this results from weak scale cross sections but is already quite large. If one is interested in strongly interacting massive particles (SIMPs), or models with very large cross sections in general, \eqref{eq:lee-weinberg} becomes exceptionally stiff, making numerical integration prohibitively difficult and quite unstable if not completely impossible. Reduction of order problems can also lead to overly optimistic error approximations, with no indication that anything is amiss. Our results do not suffer from such difficulties.

Having an analytic expression for the thermal relic density is useful in its own right, for instance in large \(\mathcal{N}\) Yang-Mills models one may be interested in the analytic behavior of thermal relic density as one takes the number of colors \(\mathcal{N}\) to infinity. This behavior can be found from \eqref{eq:results:Yinf} easily, but numerical methods must rely on extrapolation. All that is required to implement our results are standard cosmological parameters and the thermally averaged cross section as inputs, and a simple quadrature routine. The end user is not bound by the limitations of external software, thus making analysis of models that do not adhere to the typical requirements of prepackaged programs such as Lorentz invariance possible.

In addition, our method constitutes a global asymptotic approximation to the solution of a problem with an infinite order turning point. In fact, this procedure can be used to construct approximations to an entire class of problems of the form:
\begin{equation}
    \dv[2]{u}{x} - \qty[\lambda^{2} \func[2]{F}{x} e^{-2 x} + \func{P}{x}] u = 0.
\end{equation}
We have shown that the uniform WKB ansatz \eqref{eq:reg1-ansatz} allows one to extend the region of validity of the small \(x\) approximation sufficiently close to the turning point at \(x = \infty\) such that one can asymptotically match to the large \(x\) approximation. This has a large range of physics applications, including quantum mechanical scattering with a Yukawa type potential.

Our particular program could possibly generalize to a larger set of Boltzmann equations, but because our results rely on using a uniform WKB approximation we can only apply our procedure to systems that can be linearized. However, one could apply boundary-layer-analysis to obtain valid results for a multitude of Boltzmann equations.

\acknowledgements
We thank Stefano Profumo for many helpful discussions and much appreciated advice.  This work is partly supported by the U.S. Department of Energy grant number de-sc0010107.  The research of HHP was supported by Department of Energy grant number DE-FG02-04ER41286, and National Science Foundation grant number 1912719.

\bibliography{bibliography} 

\end{document}